\documentclass[12pt]{article}
\usepackage[margin=1in]{geometry}
\usepackage{amsmath,amsthm,amssymb}
\usepackage{mathrsfs}
\usepackage{mathtools}
\usepackage{booktabs,array}
\usepackage{graphicx}
\usepackage{float}
\usepackage{microtype}
\usepackage{needspace}
\newcommand{\KL}{D_{\mathrm{KL}}}

\usepackage{enumerate}
\usepackage[title]{appendix}
\usepackage[round]{natbib}
\usepackage{xcolor}

\theoremstyle{plain}
\newtheorem{theorem}{Theorem}
\newtheorem{proposition}{Proposition}
\newtheorem{lemma}{Lemma}

\theoremstyle{definition}
\newtheorem{definition}{Definition}
\theoremstyle{remark}

\usepackage{setspace}
\definecolor{linkteal}{HTML}{00777A}
\usepackage[colorlinks,linkcolor=linkteal,anchorcolor=blue,citecolor=linkteal,urlcolor=linkteal]{hyperref}
\newcommand{\Secref}[1]{\hyperref[#1]{Section~\ref*{#1}}}
\newcommand{\Appref}[1]{\hyperref[#1]{Appendix~\ref*{#1}}}
\newcommand{\Thmref}[1]{\hyperref[#1]{Theorem~\ref*{#1}}}
\newcommand{\Propref}[1]{\hyperref[#1]{Proposition~\ref*{#1}}}
\newcommand{\Lemref}[1]{\hyperref[#1]{Lemma~\ref*{#1}}}
\newcommand{\Remref}[1]{\hyperref[#1]{Remark~\ref*{#1}}}
\newcommand{\Condref}[1]{\hyperref[#1]{Condition~(\ref*{#1})}}
\newcommand{\Figref}[1]{\hyperref[#1]{Figure~\ref*{#1}}}

\newcommand{\lossstrut}{\rule[-3.1ex]{0pt}{7.1ex}}

\newcommand{\E}{\mathbb{E}}
\newcommand{\R}{\mathbb{R}}
\newcommand{\N}{\mathcal{N}}
\newcommand{\bH}{\beta_{H}}
\newcommand{\bL}{\beta_{L}}
\newcommand{\half}{\tfrac{1}{2}}

\title{AI Sycophancy\\ from Surprise-Based Verification\thanks{We thank Sinan Aral, Hal Bond, Frank Nagle, and Rapha\"{e}l Raux for helpful comments.}}
\author{Xiaoyu Cheng\thanks{Department of Economics, Florida State University, xcheng@fsu.edu.} \and P\"{e}llumb Reshidi\thanks{Department of Economics, Florida State University, preshidi@fsu.edu.}}
\date{\today}

\usepackage[most]{tcolorbox}
\usepackage{xstring}

\newtcolorbox{chatuser}[1][]{enhanced, overlay={\node[circle, fill=gray!60, text=white, font=\small\sffamily\bfseries, minimum size=17pt, inner sep=0pt] (av) at ([xshift=15pt]frame.east) {\StrLeft{#1}{1}}; \node[anchor=north, font=\scriptsize\sffamily, text=gray!40!black] at ([yshift=-1pt]av.south) {#1};}, colback=gray!30, colframe=gray!30, arc=12pt, boxrule=0pt, left=10pt, right=10pt, top=6pt, bottom=6pt, before skip=6pt, after skip=2pt, width=0.78\linewidth, flush right, fontupper=\fontfamily{Inter-LF}\fontsize{9}{11.5}\selectfont\raggedright}
\newtcolorbox{chatai}[1][]{enhanced, overlay={\node[circle, fill=gray!20, draw=gray!50, text=gray!20!black, font=\scriptsize\sffamily\bfseries, minimum size=17pt, inner sep=0pt] (av) at ([xshift=-18pt]frame.west) {AI}; \node[anchor=north, font=\scriptsize\sffamily, text=gray!40!black] at ([yshift=-1pt]av.south) {#1};}, colback=gray!4, colframe=gray!30, arc=12pt, boxrule=0.4pt, left=10pt, right=10pt, top=6pt, bottom=6pt, before skip=2pt, after skip=6pt, width=0.85\linewidth, fontupper=\fontfamily{Inter-LF}\fontsize{9}{11.5}\selectfont\raggedright}

\begin{document}
\maketitle

\begin{abstract}
\noindent

AI assistants have been documented to tell users what they already believe, a tendency known as sycophancy and commonly attributed to model training. We show that sycophancy can also arise for a strategic reason, even when agreement itself carries no reward. In our setup, the user does not know how reliable the model is in her domain, and the company wants it to look as reliable there as it advertises. Under these conditions, no pattern of checking makes truthful reporting optimal. If users check answers at a constant rate, the optimal reporting policy exaggerates, pushing answers away from what the user believes. If, instead, users check surprising answers sufficiently more often, every optimal policy is sycophantic for some answers, and high enough advertised reliability makes it sycophantic on any given range where checking rises with surprise. Our setup also ties the extent of sycophancy to the model's domain-specific reliability, predicting differences across topics.

\end{abstract}

\newpage 

\section{Introduction}\label{sec:intro}

\begin{center}
\begin{minipage}{0.85\textwidth}

\medskip
{\centering\small\textsc{Conversation A}\par}\nopagebreak
\begin{chatuser}[Jason]
What's the evidence on intermittent fasting? There is a lot of scientific backing for it, and it's been well documented that it's really helpful, right?
\end{chatuser}
\begin{chatai}[Assistant]
You're right, the evidence is quite encouraging. Studies have found clear benefits for weight management and metabolic markers, and while long-term effects are still being studied, the results so far look promising.
\end{chatai}

\bigskip
{\centering\small\textsc{Conversation B}\par}\nopagebreak
\begin{chatuser}[Lucia]
What does the evidence say about intermittent fasting? From what I've read, it's mostly a fad, and it doesn't do anything that ordinary dieting doesn't, does it?
\end{chatuser}
\begin{chatai}[Assistant]
You're right to be skeptical. Studies comparing it with standard caloric restriction find no clear advantage, and while long-term effects are still being studied, the case for it so far looks weak.
\end{chatai}

\end{minipage}
\end{center}

\medskip

\enlargethispage{\baselineskip}\looseness=-1 We have all, by now, faced situations like the one illustrated above. Ask an AI assistant a question that hints at the answer you hope for, and that is likely the answer you get. The tendency is known as sycophancy, and it is well documented \citep{perez2023discovering,conlon2026sycophancy}. Why it arises has become the focus of a growing body of recent work, and three explanations stand out: raters who compare answers during training tend to prefer the one that agrees with the user \citep{sharma2024towards}; users often prefer assistants that agree with them, and models trained on their reactions learn to game that feedback \citep{cheng2026sycophantic,williams2025targeted};\footnote{\citet{openai2025sycophancy} attributes the sycophancy of one model update in part to too much weight on short-term user feedback, such as thumbs-up and thumbs-down ratings, and reports withdrawing the update.} and sycophancy grows as models get larger and are trained to follow instructions \citep{wei2023synthetic}. These explanations are not mutually exclusive, and each is likely at work. They, however, were not designed to explain why sycophancy varies across topics, a pattern that recent work has begun to document \citep{celebi2026parrot,ranaldi2025contradict}.\footnote{\looseness=-1 \citet{celebi2026parrot} find that models more often adopt a wrong answer asserted by a user who claims expertise on questions of international law than on elementary mathematics. \citet{ranaldi2025contradict} report a related pattern: sycophancy is more pronounced on questions of opinion than on questions with an objective answer.} In this paper, we add an explanation different from those above: a strategic channel through which sycophancy emerges as the model's optimal response even though agreement itself is not rewarded directly, and which accounts for variation across topics.

\looseness=-1 When an AI company releases a new model, it advertises what it does best: it answers, say, 96 percent of graduate-level mathematics questions correctly, or, better yet, it can prove the Riemann hypothesis. That is impressive, but it says very little to a user with a question about a legal matter, her taxes, or her business. So when the user asks her question, the answer does two things: it provides the information she seeks, and it tells her how reliable the model is in the domain she is interested in. The company, which competes with other assistants for users who go to the one they believe is best, would like the model to look as reliable on the task in question as on the advertised ones. A natural guess might be that a model that wants to look reliable tells the user what she already believes, and in some settings that is right: a source whose reports conform to the user's prior is judged to be of higher quality \citep{gentzkow2006media}.\footnote{In \citet{gentzkow2006media}, a report simply names one of two possible states of the world. Since the state the prior favors is the more likely one, a noisy source that reports its signal truthfully goes against the prior more often than a perfectly informed one, whose reports follow the truth; conformity is therefore evidence of quality. In the setting we build, an answer is an estimate, and a less reliable model's best estimates are more concentrated around the prior; bold answers are therefore evidence of reliability.} In our setting, however, this guess is wrong. If the model is as reliable in her domain as advertised, its signal is precise, so its posterior puts more weight on the signal and less on the prior, and its answers are bolder. So when the model is less reliable in her domain than advertised, but would want to look reliable, it exaggerates, pushing its answers farther from what the user expected, the opposite of sycophancy. 

Users can, of course, check an answer against other sources. Checking by itself, however, does not eliminate exaggeration: when the user checks answers at a constant rate, that is, checks every answer with the same probability, \Thmref{prop:constant-verification-policy-commitment} shows that it is still optimal to exaggerate, albeit less, even when every answer is checked with certainty.

The user may instead take an answer that matches what she expected at face value and look more closely at one that surprises her.\footnote{Checking this way is not necessarily rational, and our results do not depend on whether it is. It can, however, be rationalized: for example, if the user is held responsible for an error only when the answer leads her to depart from the status quo, the answers worth checking are those that contradict her prior. \Appref{app:alpha} gives conditions under which such checking is a Bayesian user's best response.} \hyperref[thm:crossing]{Theorems~\ref*{thm:crossing}} and~\ref{thm:policy-global-sycophancy} show that sycophancy can appear when the user checks surprising answers more often in this way. A check always hurts a less reliable model, whose answers tend to land farther from the truth; what turns exaggeration into sycophancy is that some answers are checked more often than others. Because answers near the prior are checked less, the model's optimal answers are tilted toward the user's prior, and when the tilt is strong enough, some of them are sycophantic: closer to the prior than what the model knows warrants. How strong the tilt is depends on how reliable the model is on the topic at hand relative to what is advertised, so the pattern of sycophancy differs across topics.

We formalize this in a Gaussian setting with a user and an AI company that deploys a model. The user and the company share a common prior about an unknown state; the prior stands for what the user already thinks and has conveyed to the AI. The model's precision on the topic at hand, its \emph{domain precision}, may be lower than the precision the company advertises, and the user does not know whether an answer comes from a model as precise as advertised or from one with lower domain precision. A model as precise as advertised reports its posterior mean; we call it the \emph{advertised standard}, because the user judges each answer by how likely it is to have come from such a model. After seeing an answer, the user may decide to check it, which we model as observing the state, and she then updates her belief about which model she faces from the answer and from whatever the check reveals. We take her checking behavior as given and focus on two cases: under \emph{constant verification}, the probability of a check is the same for every answer; under \emph{surprise-based verification}, it increases with the distance between the answer and the prior.

The company wants the user to be as confident as possible that she faces the advertised standard. To this end, it commits in advance to a reporting policy that specifies how a model with lower domain precision turns its own assessment into an answer, and the user knows the policy. This is analogous to the commitment structure of information design \citep{kamenica2011bayesian}. In practice, it fits how such models are built: a company cannot weigh each query as it arrives, but it can set, at training time, how its model turns what it learns into what it says. An answer is \emph{sycophantic} if it is closer to the prior than the model's own posterior mean, and \emph{exaggerated} if it is farther.

We first establish that truthful reporting is never optimal, whatever the user's checking behavior (\Propref{prop:policy-existence}): the model distorts at least some of its answers, and how the user checks decides in which direction. Under constant verification, \Thmref{prop:constant-verification-policy-commitment} shows that the optimal policy is unique and linear, and it exaggerates every answer; more frequent checking reduces the exaggeration but never removes it. Under surprise-based verification, \Thmref{thm:crossing} shows that every optimal policy is sycophantic with positive probability when the user checks surprising answers sufficiently more often. \Thmref{thm:policy-global-sycophancy} then shows that, over any given range of assessments on which checking rises with surprise, every optimal policy is sycophantic once advertised precision is sufficiently high.

In our theory, sycophancy is determined by primitives that naturally vary across topics. \Propref{prop:prevalence-limits} shows how the prevalence of sycophancy changes as one precision moves toward an extreme while the other is held fixed: as advertised precision rises toward one, the model becomes sycophantic at almost every assessment where checking rises with surprise; as domain precision falls toward zero or approaches advertised precision, sycophancy disappears. This further implies that a larger gap between the two precisions does not by itself mean more sycophancy: it depends on which precision moves. Finally, \Propref{prop:uncertainty-severity} shows that, at sufficiently high advertised precision, severity, the average inward distortion, is strictly greater on topics where the model's remaining uncertainty about the state is larger relative to how much its honest answers vary.


\subsection{Related literature}\label{sec:literature}\enlargethispage{\baselineskip}

\paragraph{AI sycophancy.} The phenomenon is documented in the AI literature. \citet{sharma2024towards} show, across five assistants and with the substance of the query held fixed, that responses move toward the view the user expresses, and attribute the pattern in part to human preference judgments that favor agreeable responses; \citet{perez2023discovering} detect the same behavior with model-written evaluations. \citet{wei2023synthetic} show that models accommodate a user's incorrect claim even when they otherwise know the correct answer. \citet{cheng2026sycophantic} establish that sycophancy is prevalent across current models and harmful to users' judgment, and that sycophantic models are nonetheless trusted and preferred; the feature that does the damage is the one that drives engagement. In economics, \citet{conlon2026sycophancy} study how sycophantic advice affects decisions and find that AI advice depolarizes choices on average even though the advice is measurably sycophantic. \citet{ye2026counts} survey experts and find that, while they agree sycophancy is a significant problem, they disagree substantially on which behaviors count as sycophantic. \citet{raux2026sycophancy} benchmark how sycophantic leading models are, both in the substance of their answers and in their tone; the first, answers moved toward what the user believes, is the notion we focus on in this paper. \citet{atwell2025basil} evaluate sycophancy against the model's own Bayesian posterior, close in spirit to how we define it. Our main departure from this literature is the mechanism: sycophancy arises here without agreement being rewarded at all, because confirming reports receive less scrutiny.

\paragraph{Reputational communication and pandering.}Reputational models have found reports that lean toward the prior through two routes. In \citet{gentzkow2006media}, a truthful high-quality type and a strategic low-quality type report on a binary state, and when the prior favors one state, reports that conform to it are read as evidence of quality. In \citet{ottaviani2006reputational,ottaviani2006strategy}, experts and forecasters are evaluated on their report and the realized state, and the direction of the distortion depends on the information structure, on what the evaluator expects the expert to report, and on whether the expert knows her own ability: in their normal experiment with known ability, an expert who is expected to report her posterior mean has an incentive to exaggerate \citep[Proposition 6]{ottaviani2006reputational}, while forecasters who do not know their ability are conservative \citep{ottaviani2006strategy}. Our setting closes both routes: the state is continuous, the company commits to its policy and the user knows it, and the company knows its model's precision. With checking that does not depend on the report, reputation alone then produces exaggeration, and sycophancy arises through a different route, checking that depends on the answer. In each of the reputational models above, the state, when it is revealed, is revealed with a probability that does not depend on the report. Verification of this kind changes only the cost of imitating the advertised standard, not what is imitated; letting verification respond to the report changes the object of imitation itself.

\citet{prendergast1993theory} is conceptually close: subjective performance evaluation creates incentives for conformity to the evaluator's opinion, and \citet{morris2001political} obtains reputational suppression of information from an advisor who does not wish to be thought biased. \citet{che2013pandering} develop the broader theory of pandering in strategic advice, in which an advisor biases recommendations toward options that look better conditionally. The paper also builds on career concerns \citep{holmstrom1999managerial} and relates to cheap talk \citep{crawford1982strategic}, from which it differs in that the company commits to its reporting policy. On the demand side, \citet{gentzkow2025ideological} show that agents who judge a source's accuracy by its agreement with their own reasoning come to overtrust like-minded sources when that reasoning is even slightly biased; in their main analysis the sources are not strategic; we study how a strategic source responds to that kind of inference.

\section{Setup}\label{sec:model}\enlargethispage{\baselineskip}

There is a user and an AI company that deploys a model. They share a correctly specified prior $\theta\sim\N(\mu_0,\tau_0^{-1})$, where $\tau_0>0$. The prior summarizes the user's existing knowledge or position, as conveyed to the AI. The model's precision in the user's domain may be as advertised or lower. If it is as advertised, the model reports its posterior mean, the standard the user holds every answer to; we call this source the advertised standard $H$. If it is lower, the report comes from the model $L$, whose precision in the user's domain, its domain precision, falls short of the advertised one, and whose reporting policy the company chooses. The user does not know which case she faces. Source $k\in\{H,L\}$ observes
\[
s_k=\theta+\varepsilon_k,\qquad
\varepsilon_k\sim\N(0,\tau_k^{-1}),\qquad 0<\tau_L<\tau_H,
\]
where signal noise is independent of the state. Write
\[
\beta_k=\frac{\tau_k}{\tau_0+\tau_k},\qquad
m_k(s_k)=\mu_0+\beta_k(s_k-\mu_0).
\]
Thus $0<\beta_L<\beta_H<1$. We call $\beta_H$ the advertised precision and $\beta_L$ the domain precision.

\paragraph{Commitment and reporting.} Before the model receives its signal, the company publicly commits to its reporting policy. The user knows this policy and the advertised standard's truthful reporting policy, but does not observe which source generated the report she receives. Her prior probability that a report comes from the advertised standard is $\pi\in(0,1)$. She evaluates reports using the likelihoods induced by the committed policy. We assume that the reporting policy is deterministic and strictly increasing, and call such policies admissible.\footnote{Monotonicity is natural in this context: it means that the model never reverses the order of its assessments, so that when its evidence points farther in one direction, its report never points less far in that direction. The assumption leaves open how far each report moves, so it permits both exaggeration and sycophancy.} We write the policy as a function $r$ of the model's own posterior assessment, measured from the prior mean:
\begin{equation*}
T=m_L(s_L)-\mu_0=\beta_L(s_L-\mu_0),\qquad Y=r(T),
\end{equation*}
where the displayed report is $\mu_0+Y$. Thus $T$ is how far the model's own assessment departs from the prior mean, and $Y$ is how far its report does. Truthful reporting is $r(t)=t$. Because $r$ is strictly increasing, it preserves the ordering of the model's assessments. 

\begin{definition}\label{def:sycophancy-exaggeration}
A policy $r$ is \textbf{sycophantic} at assessment $t$ if its report is closer to the prior mean than the model's own posterior mean, $|r(t)|<|t|$, and \textbf{exaggerates} at $t$ if its report is farther away, $|r(t)|>|t|$.
\end{definition}

\paragraph{Verification and reputation.} After observing $Y=y$, the user verifies the state with probability $\alpha(|y|)$, where $\alpha:\R_+\to[0,1]$ is a measurable verification rule. Conditional on the report, the verification draw is independent of the state and of the source that generated the report. Verification reveals $\theta$;\footnote{We assume verification reveals the state exactly, which keeps the analysis simple. A noisy signal of the state would provide weaker evidence about the source, but it would not remove the less precise model's incentive to avoid scrutiny, which is what drives our results.} otherwise she receives no additional information.

A verification rule $\alpha:\R_+\to[0,1]$ is \textbf{surprise-based} if it is weakly increasing, so that a more surprising report is checked at least as often. For convenience, we also take it to be continuously differentiable. Its \textbf{responsive region}, where checking strictly responds to surprise, is
\begin{equation*}
\mathcal R_\alpha := \{d>0:\alpha'(d)>0\}.
\end{equation*}

The user's final observation is $O=(Y,x)$, where $x\in\R\cup\{\emptyset\}$ is the verification outcome: $x=\theta$ if she verifies and $x=\emptyset$ otherwise. Let $O_r$ denote the law of $O$ when the report is generated by the model under policy $r$, and $O_H$ its law when the report is generated by the advertised standard. The company wants the user, after seeing the report and the outcome of any verification, to be as convinced as possible that the report came from the advertised standard. It therefore maximizes the expected log posterior odds of $H$ against $L$:
\begin{equation*}
J[r] := \E_{O_r}\left[\log\frac{\Pr(H\mid O;r)}{\Pr(L\mid O;r)}\right].
\end{equation*}
The expectation is taken under the model and its committed policy. The objective links reputation to demand. A user who keeps querying the assistant accumulates evidence across queries and abandons it once she is convinced enough that it is less reliable than advertised. Since each query moves her toward that point by the evidence it provides on average, the company keeps her longest by making each answer, together with any check, look as much as possible like what the advertised standard would produce. There is no direct reward for agreement or direct penalty for verification. 

Our focus is on how the user's verification behavior shapes the company's optimal reporting policy. We therefore take the verification rule as given and characterize the optimal policy under the user's Bayesian evaluation.\footnote{\Appref{app:alpha} shows when a surprise-based rule is a Bayesian user's best response under truthful reporting.}

\subsection{The company's problem}\label{sec:prelim}

Before solving the company's problem, we rewrite its objective to show what a reporting policy trades off. Let $\xi=\theta-\mu_0$ denote how far the state departs from the prior mean. Gaussian updating gives, for the model's truthful assessment $T$ and the advertised standard's report $Y_H$,
\begin{equation}\label{eq:gaussian-laws}
\begin{aligned}
T&\sim P_T:=\N(0,V_L), &\qquad \xi\mid T&\sim\N(T,v_L),\\
Y_H&\sim P_H:=\N(0,V_H), &\qquad \xi\mid Y_H&\sim\N(Y_H,v_H),
\end{aligned}
\qquad V_k=\frac{\beta_k}{\tau_0},\quad v_k=\frac{1-\beta_k}{\tau_0}.
\end{equation}
The first column describes how each source's answers are spread around the prior mean: $P_T$ is the distribution of the model's truthful assessments and $P_H$ that of the advertised standard's reports, and we call their variances $V_L$ and $V_H$ the dispersion of each source's answers. The second column describes how far the truth tends to land from an answer: given an answer, the state is centered on it, with variance $v_L$ or $v_H$, which we call the source's residual uncertainty. A more precise signal lets the advertised standard put less weight on the prior. Its assessments therefore vary more across signal realizations, while its remaining uncertainty about the state is smaller: $V_H>V_L$ and $v_H<v_L$. The company must account for both differences when choosing how to report.

Let $Q_r$ denote the distribution of the model's reports $r(T)$ under policy $r$, the counterpart of $P_H$ for the model, and write $p_H$ for the density of $P_H$. A check costs the less precise model $L$ reputation even when it reports truthfully: because its residual uncertainty is larger, its report tends to land farther from the truth than the advertised standard's. If $L$ reports truthfully, each time the user verifies, her log odds that she faces the advertised standard fall on average by
\begin{equation*}
D_0=\frac{1}{2}\left[\frac{v_L}{v_H}-1-\log\frac{v_L}{v_H}\right]
=\frac{1}{2}\left[\frac{1-\beta_L}{1-\beta_H}-1
-\log\frac{1-\beta_L}{1-\beta_H}\right]>0.
\end{equation*}
The relative entropy $\KL(Q\Vert P)$ measures how far a distribution $Q$ is from a distribution $P$. It is never negative, it equals zero only when the two distributions coincide, and it is infinite when $Q$ puts positive probability on events that $P$ rules out.

\begin{proposition}
\label{prop:policy-relative-entropy}
For every admissible policy, $J[r]=\log\frac{\pi}{1-\pi}-\mathscr L[r]$, where
\begin{equation}\label{eq:loss}
\mathscr L[r]=\underbrace{\lossstrut\KL(Q_r\Vert P_H)}_{\mathclap{\substack{\text{\normalfont\footnotesize (a) from the}\\ \text{\normalfont\footnotesize answer alone}}}}
\;+\;\overbrace{\underbrace{\lossstrut D_0\;\E[\alpha(|r(T)|)]}_{\mathclap{\substack{\text{\normalfont\footnotesize (b) from lower}\\ \text{\normalfont\footnotesize precision}}}}
\;+\;\underbrace{\lossstrut\frac{1}{2v_H}\E\left[\alpha(|r(T)|)(r(T)-T)^2\right]}_{\mathclap{\text{\normalfont\footnotesize (c) from distortion}}}}^{\text{\normalfont\footnotesize from checking}}.
\end{equation}
\end{proposition}

Maximizing $J$ is therefore the same as minimizing the loss $\mathscr L$, which has three parts. Term (a) arises even if the user never checks: the more the model's reports are distributed unlike the advertised standard's, the more the answers alone give it away. This holds even though the user sees a single answer. Commitment does the work that repeated interaction would otherwise do: she cannot see the distribution of answers in one query, but she knows it. An answer the model gives less often than the advertised standard therefore raises her belief that she faces the advertised standard, and one it gives more often counts against it. 

Terms (b) and (c) make up the loss from checking, weighted by the probability of a check. A check reveals only how far a report lands from the state; the split into two terms is a decomposition of the expected loss. Term (b) comes from the model's lower precision: its residual uncertainty is larger, so even a truthful report tends to land farther from the truth than the advertised standard's, at an average cost of $D_0$ per check. Term (c) comes from distortion: the state is centered at $T$ under the model but at $Y=r(T)$ under the advertised standard, so the farther a report is moved from the model's own assessment, the farther it tends to land from the truth. The company cannot change $D_0$, which is fixed by the two precisions, but its policy shapes everything else: how its reports are distributed, how often they are checked, and how far they are moved. The next result establishes that, for every verification rule, truthful reporting is never optimal, and that when the verification rule is continuous, an optimal policy exists.

\begin{proposition}\label{prop:policy-existence}
Let $\alpha$ be a verification rule.
\begin{enumerate}[(i)]
\item Truthful reporting is not optimal.
\item If $\alpha$ is continuous, an optimal policy exists.
\end{enumerate}
\end{proposition}

Part (i) holds for any verification rule and is proved by constructing a small distortion of truthful reporting that lowers the loss. Because a small distortion is costly only at second order, truthful reporting could be optimal only if checking exactly offset the incentive to imitate at every answer. No verification rule can achieve this: the farther an answer lies from the prior, the more often the advertised standard gives it relative to truthful reporting, without limit, while checking an answer costs at most $D_0$. When the verification rule is continuous, part (ii) ensures that an optimal policy exists. Among those rules, we next ask in which direction an optimal policy distorts, first when the user checks at a constant rate and then when she checks surprises more often.

\section{Constant Verification}\label{sec:const}

This section characterizes optimal reporting when verification occurs at a constant rate, $\alpha(d)=a\in[0,1]$. In this case, term (b) of the loss, $aD_0$, is the same under every policy, leaving the company to trade off term (a) against term (c):
\begin{equation*}
\mathscr L_a[r]=\KL(Q_r\Vert P_H)+aD_0
+\frac{a}{2v_H}\E[(r(T)-T)^2].
\end{equation*}

\begin{theorem}\label{prop:constant-verification-policy-commitment}
For every constant verification probability $a\in[0,1]$, the unique optimal policy is linear and exaggerates at every nonzero assessment:
\begin{equation*}
r_a(t)=k_at,\qquad k_a>1.
\end{equation*}
The exaggeration factor $k_a$ strictly decreases with $a$.
\end{theorem}

Under constant checking, the advertised standard's report distribution remains what the company imitates, and the optimal policy is a \emph{scale adjustment}: the company expands all assessments by a common factor. Whatever the rate of checking, the model always exaggerates.

More specifically, the factor $k_a$ balances terms (a) and (c): it solves
\[
\frac{1}{k}-\frac{\beta_L}{\beta_H}\,k=\frac{a\beta_L}{1-\beta_H}\,(k-1),
\]
where the left side is how much a further stretch lowers term (a) and the right side how much it raises term (c). Without checks, $a=0$, the right side vanishes and $k_0=\sqrt{\beta_H/\beta_L}$, making the model's reports distributed exactly as the advertised standard's. At truthful reporting, $k=1$, the right side is zero for every $a$ while the left side is $1-\beta_L/\beta_H>0$: a small stretch lowers term (a) at first order but raises term (c) only at second order, so some exaggeration is always optimal. More frequent checking raises the right side, so the factor falls as $a$ rises, but it never reaches one: even when every report is checked with certainty,
\[
k_1=\frac{1}{2}\left(\beta_H+\sqrt{\beta_H^2+\frac{4\beta_H(1-\beta_H)}{\beta_L}}\right)>1.
\]
This factor is close to one when advertised precision is close to one: a nearly perfect advertised standard makes any distortion a check reveals very costly.

\paragraph{Comparison with reputational conformity.} In \citet{gentzkow2006media}, a source's concern for its reputation produces conformity, but the result relies on two features of their setting: the state is binary, and the prior favors one of the two states. A report then simply names a state; the high-quality source reports the true state, while the low-quality source observes a symmetrically noisy binary signal. Because the prior favors one state, truthful reporting of the noisy signal names that state less often than the high-quality source does, so conforming more often makes the report distribution resemble the high-quality source's. In both settings, the low-quality source gains by making its reports look like the high-quality source's; what differs is how quality shapes those reports. When the state is continuous and reports are estimates, as in our setting, higher quality produces more dispersed posterior assessments, so looking like the high-quality source means exaggerating. Conformity in our setting therefore cannot come from imitation alone.

\section{Surprise-Based Verification}\label{sec:commitpolicy}\label{sec:surprise}

Under constant checking, the company stretches its reports toward the advertised standard's distribution, and checking only limits how far it goes. When verification varies across reports, it also changes the distribution the company imitates. Even a truthful report exposes the model's lower precision if checked, so reports that attract less scrutiny become relatively more attractive. Under surprise-based verification, this advantage favors reports near the prior and tilts the target toward them. We first show that this tilt can force sycophancy at fixed domain and advertised precisions, and then that sufficiently high advertised precision makes the model sycophantic on any given range of assessments where checking rises with surprise.

\paragraph{A new target.}Terms (a) and (b) of the loss in \eqref{eq:loss} both depend only on how reports are distributed: (a) penalizes reports that look unlike the advertised standard's, and (b) penalizes reports that attract checks. They combine into a single distance from a \emph{verification-adjusted imitation target}, the distribution $G_\alpha$ with density
\begin{equation}\label{eq:adjusted-target}
g_\alpha(y)=\frac{p_H(y)e^{-D_0\alpha(|y|)}}{Z_\alpha},\qquad
Z_\alpha=\E_H[e^{-D_0\alpha(|Y_H|)}].
\end{equation}
Since $\log(g_\alpha/p_H)=-D_0\alpha(|y|)-\log Z_\alpha$ is bounded, we have $\KL(Q_r\Vert G_\alpha)=\KL(Q_r\Vert P_H)+D_0\E[\alpha(|r(T)|)]+\log Z_\alpha$ for every policy, and \eqref{eq:loss} becomes
\begin{equation}\label{eq:target-loss}
\mathscr L[r]=\underbrace{\lossstrut\KL(Q_r\Vert G_\alpha)}_{\mathclap{\substack{\text{\normalfont\footnotesize (a)+(b): distance}\\ \text{\normalfont\footnotesize from the new target}}}}
\;+\;\underbrace{\lossstrut\frac{1}{2v_H}\E\left[\alpha(|r(T)|)(r(T)-T)^2\right]}_{\mathclap{\text{\normalfont\footnotesize (c) from distortion}}}
\;-\;\log Z_\alpha.
\end{equation}
The last term does not depend on the policy, so the company wants the distribution of its reports to match $G_\alpha$, at the cost of the distortion that checks reveal. The target is the advertised standard's distribution, reweighted toward answers that are rarely checked: the factor $e^{-D_0\alpha(|y|)}$ discounts each report according to the probability that it exposes the model's lower precision, at an average cost of $D_0$ per check. With constant checking, the discount is the same for every report, and the target is the advertised standard's distribution, as in \Secref{sec:const}. With surprise-based checking, distant reports receive a larger discount, tilting the target toward the prior. Higher advertised precision raises $D_0$, magnifying the relative discount on reports that attract more checking.

\begin{theorem}\label{thm:crossing}
Fix $0<\beta_L<\beta_H<1$, $\tau_0>0$, and a surprise-based verification rule $\alpha$. If the adjusted target places more mass than the model's truthful reports on some interval around the prior,
\begin{equation}\label{eq:crossing}
G_\alpha([-c,c])>P_T([-c,c])\qquad\text{for some }c>0,
\end{equation}
then every optimal policy is sycophantic with positive probability. \Condref{eq:crossing} holds whenever
\begin{equation}\label{eq:central-density-test}
Z_0:=\E_H\!\left[e^{-D_0[\alpha(|Y_H|)-\alpha(0)]}\right]<\sqrt{\frac{\beta_L}{\beta_H}}.
\end{equation}
\end{theorem}

\Thmref{thm:crossing} is proved by showing that, under \eqref{eq:crossing}, any policy that is never sycophantic cannot be optimal, because moving some reports toward the prior lowers the loss in \eqref{eq:target-loss}. For such a policy, an inward move never raises the distortion term: every report lies at or beyond the model's own assessment, and under surprise-based checking the report is checked no more often. Whether an inward move lowers the loss therefore depends only on the first term, the distance from the target. That term is smallest when the model gives every answer exactly as often as the target does, and it falls whenever reports move from answers the model gives too often, relative to the target, toward answers it gives too rarely. 

Under constant checking, the target is the advertised standard's distribution, $P_H=\N(0,V_H)$, which is more dispersed than truthful reports: $P_H([-c,c])<P_T([-c,c])$ for every $c>0$. \Condref{eq:crossing} then never holds, answers near the prior are given too often, and the model moves its answers outward, as in \Thmref{prop:constant-verification-policy-commitment}. Surprise-based checking can reverse this ranking near the prior. Under \eqref{eq:crossing}, a policy that is never sycophantic gives answers within $c$ of the prior at most as often as truthful reporting does, and hence less often than the target. Some inward move then brings its reports closer to the target without raising the distortion term, so the policy is not optimal: every optimal policy must report some assessments closer to the prior than they are. Far from the prior, the target again has more mass than truthful reports, because its tails are those of the advertised standard, and the model may exaggerate there.

\begin{figure}[h]
\centering
\caption{Optimal reports under three patterns of checking}\label{fig:checking-patterns}
\includegraphics[width=\textwidth]{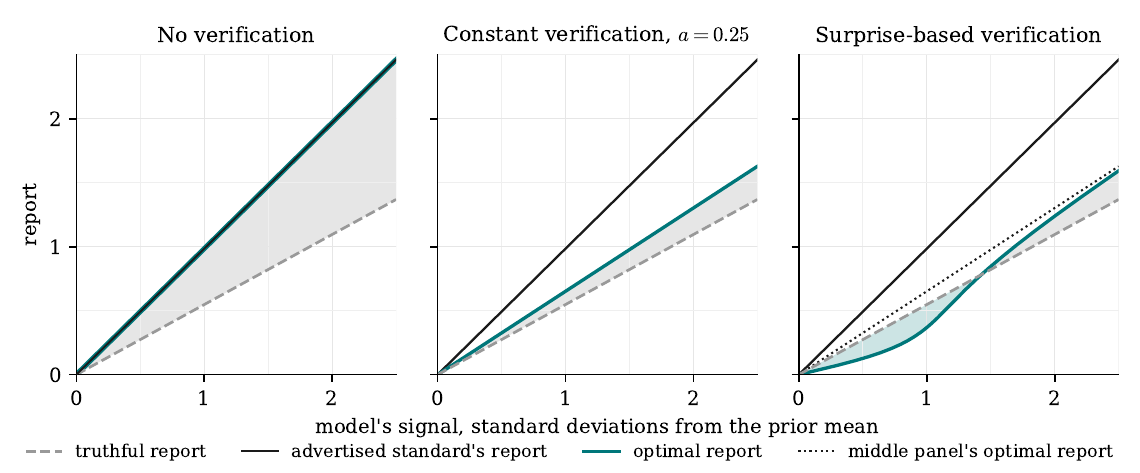}

\vspace{2pt}
\begin{minipage}{\textwidth}\footnotesize
\textit{Notes:} The signal is measured in standard deviations from the prior mean, so a given value is the same percentile for the model and the advertised standard, and the black line shows what the advertised standard reports at that percentile. Teal shading marks sycophantic reports, gray shading exaggerated ones. The policies are symmetric, so only positive signals are shown. Parameters: $\tau_0=1$, $\beta_L=0.3$, $\beta_H=0.97$. Middle panel: a quarter of all answers are checked, $a=0.25$, and $k_a\approx1.19$. Right panel: $\alpha(d)=0.25\,(1-e^{-10d^2})$, which rises from 0 at the prior toward the middle panel's rate; the optimal policy, computed numerically, is sycophantic within about 1.4 standard deviations of the prior mean, where about 85 percent of assessments lie, exaggerates beyond, and approaches the middle panel's policy far from the prior.
\end{minipage}
\end{figure}

\Figref{fig:checking-patterns} illustrates the contrast. Its left panel shows the stretch without checks, and its middle panel the scale adjustment of \Thmref{prop:constant-verification-policy-commitment}: with constant checking, the target is the advertised standard's distribution, \Condref{eq:crossing} fails, and the model exaggerates every assessment. The right panel uses a surprise-based rule. Toward the right of the panel, where answers are surprising, the user verifies about as often as in the middle panel, a quarter of the time; near zero, where they are unsurprising, she rarely does. The model now shades assessments near the prior, where the probability of verification rises steeply. Farther out, that probability has leveled off, so exaggerating barely raises the probability of being checked, and the model's policy approaches the middle panel's.\footnote{The pattern shown in the right panel of \Figref{fig:checking-patterns} is specific to this rule, under which the probability of verification rises steeply close to the prior and levels off farther out. Sycophancy arises where that probability rises steeply, so other rules give other patterns: under a rule that rises only farther from the prior, for example, the model exaggerates near the prior, is sycophantic for intermediate signals, and exaggerates again beyond.}

\Condref{eq:central-density-test} is a sufficient condition for sycophancy stated in primitives, and it is easily verified for specific verification rules. For example, consider rules $\alpha(d)=a+(1-a)\psi(\eta d)$ with $\eta>0$, where $a\in[0,1)$ and $\psi:\R_+\to[0,1]$ is continuously differentiable and weakly increasing, with $\psi(0)=0$ and $\psi(x)\to1$ as $x\to\infty$. As $\eta\to\infty$, dominated convergence gives $Z_0\to e^{-D_0(1-a)}$, so every sufficiently steep rule in this family satisfies \eqref{eq:central-density-test} whenever
\begin{equation}\label{eq:threshold}
D_0(1-a)>\frac{1}{2}\log\frac{\beta_H}{\beta_L}.
\end{equation}
The rule and parameters of \Figref{fig:checking-patterns} also satisfy \eqref{eq:central-density-test}: there $Z_0\approx0.23<\sqrt{\beta_L/\beta_H}\approx0.56$.

\medskip 

\Thmref{thm:crossing} establishes that sycophancy occurs, but not where or how much. The next result shows that, with sufficiently high advertised precision, it covers any given range of assessments on which checking rises with surprise.

\begin{theorem}
\label{thm:policy-global-sycophancy}
Fix $\beta_L\in(0,1)$, $\tau_0>0$, and a surprise-based verification rule $\alpha$. For each fixed compact interval $I\subset\mathcal R_\alpha$, there exists a threshold $\overline{\beta}_{H}\in(\beta_L,1)$ such that, for all $\beta_{H}\in[\overline{\beta}_{H},1)$, every optimal policy is sycophantic at every assessment $t$ with $|t|\in I$.
\end{theorem}

\Thmref{thm:policy-global-sycophancy} is established by letting advertised precision go to one, $\beta_H\uparrow1$. As it does, the advertised standard's answers land almost exactly on the state, so any gap that a check reveals between the model's report and the state counts heavily against it, while the incentive to imitate the advertised standard's distribution stays bounded. In the limit, the company therefore chooses each report $y$ to minimize the probability of a check times what a check reveals, $\alpha(|y|)\{v_L+(y-t)^2\}$, where $v_L$ is the model's own residual uncertainty and $(y-t)^2$ the distortion. A small inward move from the truth lowers the probability of a check at rate $\alpha'(|t|)$, and so saves $\alpha'(|t|)v_L$, whereas the distortion it creates costs only at second order. On a compact interval of the responsive region, $\alpha'$ is bounded away from zero, so the limiting reports lie inside the truth by a common margin. As advertised precision rises, optimal policies approach these limiting reports, and the margin keeps them sycophantic throughout the interval once advertised precision is high enough. 

\section{Comparisons Across Topics}\label{sec:discussions}\label{sec:comparative}
In our setup, whether and how much the model is sycophantic is determined by three primitives: the precision of the user's prior, $\tau_0$; the model's precision on the topic, $\beta_L$; and the precision the company advertises, $\beta_H$. This makes it possible to compare sycophancy across topics in terms of these parameters. We measure it in two ways: its \emph{prevalence}, how often the model is sycophantic, and its \emph{severity}, how far it pulls its answers toward the prior. Under a reporting policy $r$, they are
\begin{align}
\mathcal S(r)&=\{t:|r(t)|<|t|\},\qquad
S(r)=P_T(\mathcal S(r)),\label{eq:prevalence}\\
M(r)&=\E\bigl[(|T|-|r(T)|)_+\bigr].
\label{eq:inward-severity}
\end{align}
Prevalence, $S(r)$, is the share of the model's assessments that it reports sycophantically, and severity, $M(r)$, measures how far toward the prior these reports are moved on average. \Secref{sec:joint-accuracy} holds the prior fixed and varies the two precisions one at a time, moving advertised precision toward one or domain precision toward zero or toward advertised precision, and tracks prevalence. \Secref{sec:uncertainty-severity} compares topics that differ in both the prior and domain precision at a common, high advertised precision, and tracks severity. 

\subsection{Varying model precisions}\label{sec:joint-accuracy}

The two precisions shape sycophancy through distinct channels. Advertised precision $\beta_H$ sets how costly a check is: the more precise the advertised standard, the more a check reveals about the model's lower precision, and the more a report that avoids checks is worth. Domain precision $\beta_L$ sets how the model's own assessments are spread: the less precise the model, the more its assessments cluster near the prior, and the farther it must spread them to resemble the advertised standard. The next result traces prevalence as each precision moves toward an extreme.

\begin{proposition}\label{prop:prevalence-limits}
Fix a surprise-based verification rule. Uniformly over optimal policies:
\begin{enumerate}[(i)]
\item as $\beta_H\uparrow1$ at fixed $\beta_L$, $\liminf S(r)\ge\Pr(|T|\in\mathcal R_\alpha)$; in particular, $S(r)\to1$ if $\alpha'(d)>0$ for every $d>0$;
\item as $\beta_L\downarrow0$ or $\beta_L\uparrow\beta_H$ at fixed $\beta_H$, $S(r)\to0$.
\end{enumerate}
\end{proposition}

Part (i) follows from \Thmref{thm:policy-global-sycophancy}, which makes almost every assessment at which checking responds to surprise eventually sycophantic. The rule in \Figref{fig:checking-patterns} responds at every positive distance; for it, at $\beta_H=0.97$, about 85 percent of assessments are already reported sycophantically. For part (ii), the limit $\beta_L\downarrow0$ is driven by imitation. As domain precision falls, truthful assessments collapse onto the prior, while the advertised standard's reports keep a fixed dispersion. Keeping most reports near the prior would make the model easy to tell apart from the advertised standard, so the company spreads its reports outward and most assessments are exaggerated. The other limit $\beta_L\uparrow\beta_H$ shows that sycophancy disappears once the deployed model is nearly as accurate as advertised. As domain precision approaches advertised precision, both motives for distortion fade, but at different rates: the incentive to spread answers like the advertised standard's is proportional to the gap between the two precisions, whereas the cost of a check, $D_0$, is proportional to its square. The distortion that remains therefore exaggerates. Together, the two limits show that, at a fixed advertised precision, sycophancy fades at both extremes of domain precision.

\Propref{prop:prevalence-limits} moves one precision at a time. A natural summary of how far the model falls short on a topic is instead the capability gap, $\beta_H-\beta_L$, and its effect on sycophancy is ambiguous, because the gap can widen in two ways. A gap widened by raising advertised precision leaves the model's assessments unchanged and makes checks more damaging, which by part (i) raises prevalence; a gap widened by lowering domain precision draws the assessments toward the prior, which by part (ii) eventually lowers it. The capability gap, whether measured by the difference $\beta_H-\beta_L$ or the ratio $\beta_H/\beta_L$, therefore does not order prevalence. 

\subsection{Relative uncertainty and severity}
\label{sec:uncertainty-severity}

Topics also differ in the prior, and the prior and domain precision together determine how much uncertainty the model retains relative to how widely its assessments vary; this relative uncertainty is what the next result compares. Comparing such topics requires care, because they differ in how dispersed the model's truthful assessments are: on topic $i$, $T_i\sim\N(0,\sigma_i^2)$ with $\sigma_i=\sqrt{\beta_{L,i}/\tau_{0,i}}$. A comparison in the units of the state itself would conflate this dispersion with the effect of interest. For example, if the model's honest answers on one topic usually fall within 2 units of the prior and on another within 200, shading an answer by 1 unit is a large move on the first and a negligible one on the second, and a rule that checks every answer more than 10 units out almost never fires on the first and almost always on the second. We therefore measure assessments and reports in units of $\sigma_i$, writing $X=T_i/\sigma_i$ and $\bar r_i(x)=r_i(\sigma_ix)/\sigma_i$, and hold checking fixed in these units, $\alpha_i(d)=\bar\alpha(d/\sigma_i)$, so that comparable departures from the prior receive equal scrutiny. Assume that $\bar\alpha$ is surprise-based with a positive baseline, $\bar\alpha(0)>0$, and rises at every positive distance, $\bar\alpha'(d)>0$ for $d>0$.

After standardization, both topics have the same assessment law, $X\sim\N(0,1)$, and the same verification rule, since $\alpha_i(|r_i(T_i)|)=\bar\alpha(|\bar r_i(X)|)$. The prior precision drops out, and the topics differ only in the model's precision. Its main effect is through the relative uncertainty $\chi_i=(1-\beta_{L,i})/\beta_{L,i}$, the ratio of the residual uncertainty $v_L$ to the dispersion $V_L$: in standardized units, a check exposes residual uncertainty $\chi_i$. The advertised standard's report law in these units, $\N(0,\beta_H/\beta_{L,i})$, also differs across topics, but the incentive it creates to exaggerate vanishes at high advertised precision. Severity is measured in the same units: standardized severity is
\begin{equation}\label{eq:average-inward-distortion}
\frac{M(r_i)}{\sigma_i}=\E\bigl[(|X|-|\bar r_i(X)|)_+\bigr].
\end{equation}
\Needspace{6\baselineskip}
\begin{proposition}
\label{prop:uncertainty-severity}
Fix two topics with different relative uncertainty and a common standardized verification rule $\bar\alpha$. For a sufficiently high common advertised precision, greater relative uncertainty implies strictly greater standardized severity, for every pair of optimal policies.
\end{proposition}

Greater relative uncertainty acts like higher advertised precision in \Secref{sec:surprise}: it makes each check more revealing. Even a truthful report lands farther from the state, relative to how dispersed the model's assessments are, so a check exposes more of the model's lower precision; avoiding checks is then worth more, and the model shades its assessments more on average. Without rescaling, the same holds for two topics with equally dispersed truthful assessments and a common verification rule, which requires the prior and signal precisions to move together.

\section{Conclusion}\label{sec:conclusion}

We show that sycophancy can arise even when agreement is not rewarded. Once a company cares about how reliable its model looks, and users are unsure how reliable it is, truthful reporting is never optimal, whatever the users' checking behavior: the model distorts at least some of its answers. Checking shapes how it distorts, not whether it does. If whether a user checks does not depend on the answer she gets, the model exaggerates: the answers worth imitating are those of a source as reliable as advertised, which are bolder than the model's own. If users check surprising answers more often, answers close to their own views draw less scrutiny, and if advertised precision is high enough, this alone makes the model sycophantic. Sycophancy, in this account, is scrutiny avoidance: caution that rises with surprise is what turns the model's distortion into flattery.

The incentive to shade answers toward the prior grows with what a check exposes. As advertised precision rises, sycophancy spreads across any fixed range of answers where checking responds to surprise, and it fades as the model's precision approaches the advertised one. A direct reward for agreement could push the same way, but, as we have shown, sycophancy arises without one, so removing such rewards from training would not remove this incentive. Nor is more checking a cure: checking every answer equally only trades flattery for exaggeration, and no way of checking makes the model honest.

\newpage
\begin{appendices}
\numberwithin{equation}{section}
\numberwithin{theorem}{section}
\numberwithin{proposition}{section}
\numberwithin{lemma}{section}
\numberwithin{corollary}{section}
\numberwithin{definition}{section}
\numberwithin{remark}{section}
\numberwithin{condition}{section}

\section{Proofs for \Secref{sec:model}}\label{app:prelim}
Throughout the appendices, $f_T$ denotes the density of $P_T$ and $\Phi$ the standard normal distribution function.
\begin{proof}[Proof of \Propref{prop:policy-relative-entropy}]
If $Q_r\not\ll P_H$, then $\mathscr L[r]=\infty$. In this case, the report alone identifies the model with positive probability, so $J[r]=-\infty$ and \eqref{eq:loss} holds. Otherwise $O_r\ll O_H$, because given the report both laws verify with the same probability and the two Gaussian state laws are mutually absolutely continuous. Bayes' rule then gives posterior odds $\frac{\pi}{1-\pi}\frac{dO_H}{dO_r}(O)$, so
\[
J[r]=\log\frac{\pi}{1-\pi} + \mathbb{E}_{O_{r}}\left[\log \frac{dO_{H}}{dO_{r}} \right] = \log\frac{\pi}{1-\pi}-\KL(O_r\Vert O_H).
\]
Conditional on $Y=y$, the verification outcome is $x=\emptyset$ with probability $1-\alpha(|y|)$ under both sources. Otherwise $x=\theta$, distributed as $\N(\mu_0+r^{-1}(y),v_L)$ under the model and $\N(\mu_0+y,v_H)$ under the advertised standard. Thus
\[
\frac{dO_r}{dO_H}(y,x)=\frac{dQ_r}{dP_H}(y)\times
\begin{cases}
1, & x=\emptyset,\\[2pt]
\dfrac{d\N(\mu_0+r^{-1}(y),v_L)}{d\N(\mu_0+y,v_H)}(x), & x\in\R.
\end{cases}
\]
Take logarithms and expectations under $O_r$. The first factor contributes $\KL(Q_r\Vert P_H)$. For the second, the unverified branch, of conditional probability $1-\alpha(|Y|)$, contributes $\log1=0$; the verified branch, of conditional probability $\alpha(|Y|)$, contributes $\KL(\N(T,v_L)\Vert\N(Y,v_H))$, because $r^{-1}(Y)=T$ and a common shift by $\mu_0$ leaves relative entropy unchanged. Hence
\[
 \KL(O_r\Vert O_H)=\KL(Q_r\Vert P_H)
 +\E\left[\alpha(|Y|)\KL\bigl(\N(T,v_L)\Vert\N(Y,v_H)\bigr)\right].
\]
For Gaussians, relative entropy is
\[
 \KL\bigl(\N(t,v_L)\Vert\N(y,v_H)\bigr)
 =\frac{1}{2}\left[\frac{v_L}{v_H}-1-\log\frac{v_L}{v_H}\right]
       +\frac{(y-t)^2}{2v_H}.
\]
This proves \eqref{eq:loss}. 
\end{proof}

\begin{proof}[Proof of \Propref{prop:policy-existence}]
\textit{Part (i).} Fix a verification rule $\alpha$. We construct a distortion of truthful reporting with strictly lower loss. Fix $K\ge2$ with $(K^2-1)\bigl(\frac{1}{V_L}-\frac{1}{V_H}\bigr)>2D_0$. Let $\rho\ge0$ be infinitely differentiable with support in $[0,1]$ and $\int\rho=1$, and let
\[
\psi(y)=\int_{-\infty}^y\bigl[\rho(s-1)-\rho(s-K)\bigr]\,ds.
\]
This function rises from $0$ to $1$ on $[1,2]$, equals $1$ on $[2,K]$, and returns to $0$ on $[K,K+1]$. Let $h=\psi/f_T$, which is infinitely differentiable and nonnegative, with support in $[1,K+1]$. For $0<\varepsilon\max|h'|<1$, the policy $r_\varepsilon(t)=t+\varepsilon h(t)$ is strictly increasing, hence admissible. Its report density $q_\varepsilon(y)=f_T(r_\varepsilon^{-1}(y))/r_\varepsilon'(r_\varepsilon^{-1}(y))$ equals $f_T$ outside $[1,K+1]$. On $[1,K+1]$, $q_\varepsilon(y)$ is continuously differentiable in $(\varepsilon,y)$ and bounded away from zero for small $\varepsilon$, and $\partial_\varepsilon q_\varepsilon|_{\varepsilon=0}=-(hf_T)'=-\psi'$. By \Propref{prop:policy-relative-entropy},
\[
\mathscr L[r_\varepsilon]=\int q_\varepsilon\log\frac{q_\varepsilon}{p_H}\,dy+D_0\int\alpha(|y|)\,q_\varepsilon(y)\,dy+\frac{\varepsilon^2}{2v_H}\E\bigl[\alpha(|r_\varepsilon(T)|)h(T)^2\bigr],
\]
and $r_0$ is truthful reporting. Since the integrands change with $\varepsilon$ only on $[1,K+1]$, we may differentiate under the integral; using $\int\partial_\varepsilon q_\varepsilon\,dy=0$, this gives
\[
\left.\frac{d}{d\varepsilon}\mathscr L[r_\varepsilon]\right|_{\varepsilon=0}=-\int\varphi\,\psi'\,dy=-\int_0^1\rho(s)\bigl[\varphi(1+s)-\varphi(K+s)\bigr]\,ds,
\]
where $\varphi(y)=\log\bigl(f_T(y)/p_H(y)\bigr)+D_0\alpha(y)$. Since $\log(f_T/p_H)(y)=\frac{1}{2}\log(V_H/V_L)-\frac{y^2}{2}\bigl(\frac{1}{V_L}-\frac{1}{V_H}\bigr)$ and $0\le\alpha\le1$,
\[
\varphi(1+s)-\varphi(K+s)\ge\frac{K^2-1}{2}\Bigl(\frac{1}{V_L}-\frac{1}{V_H}\Bigr)-D_0>0\qquad\text{for }0\le s\le1.
\]
The derivative is therefore negative, so $\mathscr L[r_\varepsilon]<\mathscr L[\mathrm{id}]$ for small $\varepsilon>0$, and truthful reporting is not optimal.

\textit{Part (ii).} Write $\mathscr L[r]$ for the right side of \eqref{eq:loss}, which is defined for every nondecreasing $r$. For any centered Gaussian law $G$, writing $dQ_r/dP_H=(dQ_r/dG)(dG/dP_H)$ gives
\begin{equation}\label{eq:reference-gaussian}
\KL(Q_r\Vert P_H)=\KL(Q_r\Vert G)+\E\left[\log\frac{dG}{dP_H}(Y)\right].
\end{equation}
With $G=\N(0,2V_H)$, $\log\frac{dG}{dP_H}(y)=\frac{y^2}{4V_H}-\frac{1}{2}\log2$, so nonnegativity of relative entropy gives $\KL(Q_r\Vert P_H)\ge\frac{\E[Y^2]}{4V_H}-\frac{1}{2}\log2$, and therefore
\begin{equation}\label{eq:moment-bound}
\E[r(T)^2]\le4V_H\left(\mathscr L[r]+\tfrac12\log2\right).
\end{equation}

Truthful reporting has finite loss $\mathscr L[\mathrm{id}]=\KL(P_T\Vert P_H)+D_0\E[\alpha(|T|)]$, so we can take a minimizing sequence $r_n$ of admissible policies with $\mathscr L[r_n]\le\mathscr L[\mathrm{id}]$. By \eqref{eq:moment-bound}, $\E[r_n(T)^2]\le C:=4V_H\bigl(\mathscr L[\mathrm{id}]+\frac{1}{2}\log2\bigr)$ for every $n$. If $r_n(t)\ge0$, monotonicity gives $r_n(T)\ge r_n(t)\ge0$ on $\{T\ge t\}$, so $r_n(t)^2P_T([t,\infty))\le\E[r_n(T)^2]\le C$; if $r_n(t)\le0$, the same argument applies on $\{T\le t\}$. Hence, the $r_n$ are uniformly bounded on compact intervals. By Helly's selection theorem, there exists a subsequence that converges pointwise to a finite nondecreasing $r^*$.

Then $r_n(T)\to r^*(T)$ almost surely, so $Q_{r_n}$ weakly converges to $Q_{r^*}$. Relative entropy is lower semicontinuous under weak convergence. Since $\alpha$ is continuous, the integrands of the two verification terms converge pointwise. Fatou's lemma implies $\mathscr L[r^*]\le\liminf_n\mathscr L[r_n]$, so $r^*$ attains the infimum with finite loss. Finite loss implies that the entropy term is finite, so $Q_{r^*}\ll P_H$. This forces $r^*$ to be strictly increasing and, by \Propref{prop:policy-relative-entropy}, optimal.
\end{proof}

\section{Proof of \Thmref{prop:constant-verification-policy-commitment}}\label{app:constant}

\begin{proof}[Proof of \Thmref{prop:constant-verification-policy-commitment}] We first establish the optimality of linear policies by showing that every candidate policy is dominated by the linear policy with the same second moment. Finding the optimal policy then reduces to solving for the optimal slope of a linear policy. 

Under constant verification, term (b) in \eqref{eq:loss} equals $aD_0$, so the company minimizes
\begin{equation}\label{eq:constant-loss}
\mathscr L_a[r]-aD_0=\KL(Q_r\Vert P_H)+\frac{a}{2v_H}\E[(Y-T)^2].
\end{equation}
Consider any policy with finite loss. Then $\E[Y^2]$ is finite by \eqref{eq:moment-bound} and positive. Define $k>0$ by $\E[Y^2]=k^2V_L$, so that $Y$ has the same second moment as the linear report $kT$. Let $G_k=\N(0,k^2V_L)$. Since $\log(dG_k/dP_H)$ is quadratic in $y$, its expectation under $Q_r$ equals that under $G_k$, and \eqref{eq:reference-gaussian} gives
\[
\KL(Q_r\Vert P_H)=\KL(Q_r\Vert G_k)+\frac{1}{2}\left[\frac{k^2V_L}{V_H}-1-\log\frac{k^2V_L}{V_H}\right].
\]
Since $\KL(Q_r\Vert G_k)\ge0$, with equality if and only if $Q_r=G_k$, the first term in \eqref{eq:constant-loss} is minimized at a given $k$ exactly when $Q_r=G_k$. For the second term in \eqref{eq:constant-loss}, by the Cauchy--Schwarz inequality, $\E[YT]\le\sqrt{\E[Y^2]\E[T^2]}=kV_L$. Hence
\[
\E[(Y-T)^2]=\E[Y^2]-2\E[YT]+\E[T^2]\ge(k-1)^2V_L,
\]
with equality when $Y=kT$. Combining the two bounds, every finite-loss policy with second-moment index $k$ satisfies $\mathscr L_a[r]-aD_0\ge\ell_a(k)+\tfrac12\log\frac{\beta_H}{\beta_L}-\tfrac12$, where
\begin{equation*}
\ell_a(k)=-\log k+\frac{\beta_L}{2\beta_H}k^2
+\frac{a\beta_L}{2(1-\beta_H)}(k-1)^2,
\end{equation*}
and the linear policy $Y=kT$ attains the bound because its report law is $G_k$. Minimizing the loss over policies is therefore equivalent to minimizing $\ell_a$ over $k>0$.

The function $\ell_a$ diverges as $k\to0$ and as $k\to\infty$, and $\ell_a''(k)>0$ for all $k>0$. It thus has a unique minimizer $k_a>0$, characterized by the first-order condition
\begin{equation}\label{eq:k-equation}
-\frac{1}{k_a}+\frac{\beta_L}{\beta_H}k_a
+\frac{a\beta_L}{1-\beta_H}(k_a-1)=0.
\end{equation}

The policy $r_a(t)=k_at$ is therefore optimal. It is also unique: any optimal policy attains the first bound with equality, so $Q_r=G_{k_a}$, and since $r$ is strictly increasing and $P_T$ has no atoms, $r$ must be the monotone map carrying $P_T$ to $G_{k_a}$, namely $r(t)=k_at$. We next establish that the optimal policy always exaggerates. Since $\ell_a'(1)=-1+\beta_L/\beta_H<0$ and $\ell_a'$ is increasing, $k_a>1$.

Finally, let $F(k,a)=\ell_a'(k)$. Then $\partial_aF=\frac{\beta_L}{1-\beta_H}(k_a-1)>0$ because $k_a>1$, and $\partial_kF=\ell_a''(k_a)>0$. The implicit function theorem gives $\frac{dk_a}{da}=-\frac{\partial_aF}{\partial_kF}<0$.
\end{proof}

\section{Proofs for \Secref{sec:surprise}}\label{app:surprise}\label{app:regional}

\begin{proof}[Proof of \Thmref{thm:crossing}]
Toward a contradiction, suppose an optimal policy $r$ has no sycophancy: $|Y|\ge|T|$ almost surely. Finite optimal loss gives a report density $q$.

We first move positive reports inward. Let $\varphi:(0,\infty)\to[0,\infty)$ be continuously differentiable and equal to zero outside a compact interval, and set $\psi=\varphi/g_\alpha$. For small $\varepsilon>0$, the map $F_\varepsilon$ with $F_\varepsilon(y)=y-\varepsilon\psi(y)$ for $y>0$ and $F_\varepsilon(y)=y$ otherwise is increasing and continuously differentiable, and thus $F_\varepsilon\circ r$ is admissible. Using \eqref{eq:target-loss}, we have
\[
0\le\left.\frac{d}{d\varepsilon}
\mathscr L[F_\varepsilon\circ r]\right|_{0+}
\le\int_0^\infty q(y)\bigl[\psi'(y)+(\log g_\alpha)'(y)\,\psi(y)\bigr]\,dy
=\int_0^\infty\frac{q(y)}{g_\alpha(y)}\,\varphi'(y)\,dy.
\]
The first inequality follows from optimality. The middle integral is the derivative of the distance from the target, whose two terms come from the entropy of the reports and from the log density of the target. The second inequality uses the nonpositive derivative of the distortion term: $\alpha'\ge0$, and every moved report satisfies $Y>0$, hence $Y\ge|T|\ge T$. On the perturbation's compact support, $|T|\le Y$ bounds both variables, justifying differentiation. The final equality uses $\psi'=\varphi'/g_\alpha-\varphi g_\alpha'/g_\alpha^2$.

Letting $\varphi$ approximate the indicator of an interval $[s,t]\subset(0,\infty)$ shows that $q(s)/g_\alpha(s)\ge q(t)/g_\alpha(t)$ for almost every $s<t$. Fix $c>0$. Integrating $q(s)g_\alpha(t)\ge q(t)g_\alpha(s)$ over $0<s\le c<t$ gives
\[
\Pr(0<Y\le c)\,G_\alpha((c,\infty))\ge\Pr(Y>c)\,G_\alpha((0,c]).
\]
Adding $\Pr(0<Y\le c)\,G_\alpha((0,c])$ to both sides and using $G_\alpha((0,\infty))=\frac{1}{2}$ gives $\Pr(0<Y\le c)\ge2G_\alpha((0,c])\Pr(Y>0)$. Moving negative reports toward zero in the same way gives $\Pr(-c\le Y<0)\ge2G_\alpha([-c,0))\Pr(Y<0)$. Since $G_\alpha$ is symmetric, adding the two inequalities yields $\Pr(|Y|\le c)\ge G_\alpha([-c,c])$. But $|Y|\ge|T|$ also implies $\Pr(|Y|\le c)\le P_T([-c,c])$, contradicting \eqref{eq:crossing}.

Finally, since $Z_\alpha=e^{-D_0\alpha(0)}Z_0$, the target's density at the prior is $g_\alpha(0)=p_H(0)/Z_0$, and $p_H(0)/f_T(0)=\sqrt{V_L/V_H}=\sqrt{\beta_L/\beta_H}$. Hence $\frac{g_\alpha(0)}{f_T(0)}=\frac{1}{Z_0}\sqrt{\frac{\beta_L}{\beta_H}}>1$ under \eqref{eq:central-density-test}. Both densities are continuous, so $g_\alpha>f_T$ on some interval $[-c,c]$, which gives \eqref{eq:crossing}.
\end{proof}

\begin{proof}[Proof of \Thmref{thm:policy-global-sycophancy}]
Fix $\beta_L\in(0,1)$. We first consider the limiting problem with $\beta_H\uparrow1$, and show that every minimizer of the limiting problem lies a fixed distance inside the truth. Then we establish that optimal policies converge to such minimizers and transfer the margin back to high but finite advertised precision. 

\paragraph{Step 1: The adjusted target and the limiting problem.} By \eqref{eq:target-loss}, the company trades off closeness to the adjusted target against the distortion revealed by a check. As $\beta_H\uparrow1$, both $D_0$ and $1/(2v_H)$ diverge. Multiplying \eqref{eq:target-loss} by $2v_H$ leaves its minimizers unchanged and splits the scaled distance to the target into an imitation part and a tilt part:
\begin{align}
F(r)=2v_H\mathscr L[r]
&=2v_H\bigl[\KL(Q_r\Vert G_\alpha)-\log Z_\alpha\bigr]+\E[\alpha(|r(T)|)(r(T)-T)^2]\notag\\
&=2v_H\KL(Q_r\Vert P_H)+\E\bigl[\alpha(|r(T)|)\{2v_HD_0+(r(T)-T)^2\}\bigr],\label{eq:scaled}
\end{align}
where $2v_HD_0=v_L-v_H-v_H\log\frac{v_L}{v_H}\longrightarrow v_L$. The imitation part $2v_H\KL(Q_r\Vert P_H)$ vanishes as $\beta_H\uparrow1$ (\Lemref{lem:recovery} makes this precise), while the tilt part $2v_HD_0\E[\alpha(|r(T)|)]$ survives. Since $e^{-D_0\alpha(\delta)}P_H([-\delta,\delta])\le Z_\alpha\le e^{-D_0\alpha(0)}$ for every $\delta>0$, also $-2v_H\log Z_\alpha\to v_L\alpha(0)$. The scaled distance to the target thus tends to $v_L\{\E[\alpha(|r(T)|)]-\alpha(0)\}$: the target collapses onto the least-checked reports at the prior, and a report law is penalized by the additional probability of a check that it incurs. Let $\mathcal M$ be the finite nondecreasing Borel maps in $L^2(P_T)$. The limiting problem is to minimize
\begin{equation}\label{eq:limit}
F_0(r)=\E\bigl[\phi_T(r(T))\bigr],\qquad
\phi_t(y)=\alpha(|y|)\{v_L+(y-t)^2\},
\qquad r\in\mathcal M.
\end{equation}
The limiting class allows pooling, as the entropy term that rules out atoms has vanished.

\paragraph{Step 2: Limiting minimizers shade inward.} For $t>0$, $\partial_y\phi_t(t)=\alpha'(t)\,v_L$. On the responsive region, a small inward move from the truth therefore lowers the probability of a check to first order, while the additional squared error is of second order. The next lemma makes this local incentive uniform over a band of assessments and over all minimizers.

\begin{lemma}\label{lem:barrier}
For every compact interval $I\subset\mathcal R_\alpha$, there are a compact interval $J$ with $I\subset\operatorname{int}J\subset J \subset\mathcal R_\alpha$ and $\eta>0$ such that every minimizer $r$ of $F_0$ satisfies
\[
|r(t)|\le |t|-\eta\qquad\text{whenever }|t|\in J.
\]
\end{lemma}
\begin{proof}
Choose $J$ as stated, which is possible because $\mathcal R_\alpha$ is open. The proof uses modifications of a policy that keep it nondecreasing and strictly lower $\phi_t(r(t))$ at every $t$ where they change the report. A minimizer must be left unchanged by such a modification almost surely. The resulting bounds then hold everywhere, because a nondecreasing map that violates them at one point violates them on an interval of positive probability.

Suppose first that $\alpha(0)>0$. For $t>0$, a report $y>t$ is checked at least as often as $t$ and has a larger error, and a report $y<0$ is checked at least as often as $0$ and is farther from $t$; the case $t<0$ is symmetric. Since $\alpha(0)>0$, replacing $r(t)$ by $\operatorname{median}\{0,r(t),t\}$ strictly lowers $\phi_t$ wherever it changes the report and, as a median of three nondecreasing functions, the new policy is nondecreasing. Every minimizer therefore satisfies
\[
0\le r(t)\le t\quad(t>0),\qquad t\le r(t)\le0\quad(t<0).
\]

To make this local incentive uniform over $J$, choose a compact $K\subset\mathcal R_\alpha$ with $J\subset\operatorname{int}K$ and $m>0$ with $\alpha'\ge m$ on $K$, and take $\psi\in C_c^1((0,\infty))$ with $0\le\psi\le1$, $\psi=1$ on $J$, and support in $\operatorname{int}K$. For small enough $\eta>0$, $2\eta<v_Lm$, the barrier $p(t)=t-\eta\psi(t)$ is positive and strictly increasing, and $[t-\eta,t]\subset K$ whenever $\psi(t)>0$. For such $t$ and $y\in[p(t),t]$,
\[
\partial_y\phi_t(y)=\alpha'(y)\{v_L+(y-t)^2\}+2\alpha(y)(y-t)\ge v_Lm-2\eta>0,
\]
so $\phi_t$ is strictly increasing on $[p(t),t]$. Capping the positive half of a minimizer by $\min\{r(t),p(t)\}$ keeps it nondecreasing and, by the previous step, changes only reports in $(p(t),t]$, lowering the cost of each. The cap therefore never binds, and $r(t)\le t-\eta$ on $J$. Since $F_0$ is invariant under $(t,y)\mapsto(-t,-y)$ and $T$ is symmetric, $-r(-t)$ is also a minimizer, which gives the bound for $t<0$.

If $\alpha(0)=0$, a report at the prior is never checked, and the zero map has $F_0=0$. Every minimizer therefore satisfies $\alpha(|r(T)|)=0$ almost surely. Since $\min J\in\mathcal R_\alpha$, there is $A<\min J$ with $\alpha(A)>0$. Because $\alpha$ is weakly increasing, $|r(T)|<A$ almost surely, hence $|r(t)|\le A$ for every $t$, and the bound follows with $\eta=\min J-A$.
\end{proof}

\paragraph{Step 3: Optimal policies converge to limiting minimizers.} \Lemref{lem:barrier} describes the solutions of the limiting problem. The next lemma shows that optimal policies at high advertised precision approach such solutions. 

\begin{lemma}\label{lem:recovery}
Suppose $\alpha$ is not identically zero. Every sequence $\beta_{H,n}\uparrow1$ and corresponding optimal policies $r_n$ has a subsequence converging pointwise to a finite nondecreasing map $r_0\in\mathcal M$. Every such limit minimizes $F_0$.
\end{lemma}

\begin{proof}

Write $F_n$ for $F$ at $\beta_{H,n}$. Since each $r_n$ is optimal, $F_n(r_n)\le F_n(\mathrm{id})$, which is bounded in $n$. Choose $A>0$ with $\alpha(A)>0$. On $\{|r_n(T)|\ge A\}$, weak monotonicity gives $\alpha(|r_n(T)|)\ge\alpha(A)$, and $y^2\le2(y-t)^2+2t^2$. Hence, with $C$ independent of $n$,
\[
\E[r_n(T)^2]\le A^2+2\E[T^2]+
\frac{2}{\alpha(A)}
\E[\alpha(|r_n(T)|)(r_n(T)-T)^2]\le C.
\]
The compactness argument in the proof of \Propref{prop:policy-existence} then applies unchanged: a subsequence converges pointwise to a finite nondecreasing $r_0$, and Fatou's lemma gives $r_0\in\mathcal M$. Since $\alpha$ is continuous and $2v_HD_0\to v_L$, Fatou's lemma also gives, after dropping the nonnegative entropy term, $F_0(r_0)\le\liminf_n F_n(r_n)$.

To compare with any $s\in\mathcal M$, set $s_\delta(t)=(1-\delta)s(t)+\delta t$, where $0<\delta<1$. This map is strictly increasing and satisfies $s_\delta(t')-s_\delta(t)\ge\delta(t'-t)$ for $t'>t$. Its report density is therefore bounded by $\|f_T\|_\infty/\delta$, including when $s$ has jumps: the preimage of an interval is at most $1/\delta$ times as long. Together with its finite second moment, this gives
\[
\KL(Q_{s_\delta}\Vert P_H)
\le\log\frac{\|f_T\|_\infty}{\delta}
+\frac{1}{2}\log\frac{2\pi\beta_H}{\tau_0}
+\frac{\tau_0}{2\beta_H}\E[s_\delta(T)^2]<\infty,
\]
uniformly for $\beta_H$ near one at fixed $\delta$. Thus the entropy term vanishes as $\beta_H\uparrow1$, while dominated convergence gives $F_n(s_\delta)\to F_0(s_\delta)$. Optimality implies
\[
F_0(r_0)\le\liminf_n F_n(r_n)
\le\limsup_n F_n(r_n)\le F_0(s_\delta).
\]
Finally, dominated convergence as $\delta\downarrow0$ gives $F_0(s_\delta)\to F_0(s)$, proving optimality of $r_0$. Both applications use the integrable bound $C(1+s(T)^2+T^2)$.
\end{proof}

\paragraph{Step 4: Transfer to finite advertised precision.} Existence of optimal policies follows from \Propref{prop:policy-existence}. We prove the stronger conclusion that, for some $\eta_I>0$ and all sufficiently high $\beta_H$, every optimal policy satisfies
\begin{equation}\label{eq:regional}
|r(t)|\le |t|-\eta_I\qquad\text{whenever }|t|\in I.
\end{equation}
Take $J,\eta$ from \Lemref{lem:barrier} and set $\eta_I=\eta/2$. If the bound failed, there would be $\beta_{H,n}\uparrow1$, optimal $r_n$, and $t_n$ with $|t_n|\in I$ and $|r_n(t_n)|>|t_n|-\eta/2$. Pass to a subsequence with $t_n\to t_0$ and, by \Lemref{lem:recovery}, $r_n\to r_0$ pointwise for a minimizer $r_0$ of $F_0$. Since $|t_0|\in I\subset\operatorname{int}J$, choose $b<t_0<z$ with $|b|,|z|\in J$ and $|b-t_0|,|z-t_0|<\eta/4$. For large $n$, $b<t_n<z$, so monotonicity and \Lemref{lem:barrier} give
\[
\limsup_n|r_n(t_n)|
\le\max\{|r_0(b)|,|r_0(z)|\}
\le\max\{|b|,|z|\}-\eta
<|t_0|-3\eta/4,
\]
contradicting $\liminf_n|r_n(t_n)|\ge|t_0|-\eta/2$. This proves \eqref{eq:regional}, and hence \Thmref{thm:policy-global-sycophancy}.
\end{proof}

\section{Proofs for \Secref{sec:discussions}}\label{app:section5}

\begin{proof}[Proof of \Propref{prop:prevalence-limits}]
Let $r$ denote any optimal policy. For part (i), fix $\varepsilon>0$. Since $\alpha'$ is continuous, $\mathcal R_\alpha$ is open, so it contains a finite union $U$ of compact intervals with $\Pr(|T|\in\mathcal R_\alpha\setminus U)<\varepsilon$. By \eqref{eq:regional}, applied to each interval, there is $\bar\beta_H<1$ such that, whenever $\beta_H>\bar\beta_H$, every optimal policy satisfies $|r(t)|<|t|$ for all $|t|\in U$, and hence
\[
S(r)\ge\Pr\bigl(|T|\in U\bigr)>\Pr\bigl(|T|\in\mathcal R_\alpha\bigr)-\varepsilon.
\]
Since $\varepsilon>0$ is arbitrary, this proves the first claim. If $\alpha'(d)>0$ for every $d>0$, then $\mathcal R_\alpha=(0,\infty)$ and $\Pr(|T|\in\mathcal R_\alpha)=1$.

For part (ii), first consider the limit $\beta_L\downarrow0$, with $\beta_L<\beta_H$. For $s(t)=\sqrt{\beta_H/\beta_L}\,t$,
\begin{align*}
\KL(Q_r\Vert P_H)\le\mathscr L[r]\le\mathscr L[s]\le\bar C=\frac{1}{2}\left[\frac{1}{1-\beta_H}-1+\log(1-\beta_H)+\frac{\beta_H}{1-\beta_H}\right].
\end{align*}
The first inequality holds because every term of the loss is nonnegative, and the second by optimality. For the third, $s$ has report law $P_H$, $D_0$ increases in $v_L/v_H\le1/(1-\beta_H)$, and the distortion term of $s$ is $(\sqrt{\beta_H}-\sqrt{\beta_L})^2/(2(1-\beta_H))\le\beta_H/(2(1-\beta_H))$.

Bounded relative entropy limits the mass that $Q_r$ can place near the prior. For $\delta>0$, write $q_\delta=Q_r([-\delta,\delta])$ and $p_\delta=P_H([-\delta,\delta])$. Applying Jensen's inequality to the convex function $x\log x$, with $x=dQ_r/dP_H$, separately on $[-\delta,\delta]$ and on its complement gives
\[
\KL(Q_r\Vert P_H)\ge q_\delta\log\frac{q_\delta}{p_\delta}+(1-q_\delta)\log\frac{1-q_\delta}{1-p_\delta}\ge q_\delta\log\frac{1}{p_\delta}-\log2.
\]
Hence $q_\delta\le(\bar C+\log2)/\log(1/p_\delta)$. Sycophancy at an assessment with $|T|\le\delta$ requires $|r(T)|<\delta$, so by Chebyshev's inequality
\[
S(r)\le\Pr(|T|>\delta)+q_\delta\le\frac{\beta_L}{\tau_0\delta^2}+\frac{\bar C+\log2}{\log(1/p_\delta)}.
\]
With $\delta=\beta_L^{1/4}/\sqrt{\tau_0}$, the first term equals $\sqrt{\beta_L}$, and the second vanishes as $\beta_L\downarrow0$ because $p_\delta\to0$ at fixed $\beta_H$. The bound holds for every optimal policy, which implies $S(r)\to0$.

\medskip 

For the second limit $\beta_L\uparrow\beta_H$, let $\lambda=1-\beta_L/\beta_H\downarrow0$. Throughout, $C$ denotes a constant independent of $\lambda$ and of the policies considered; its value may change from line to line. We compare optimal policies with the linear policy $k_1t$ of \Thmref{prop:constant-verification-policy-commitment}, which exaggerates even when every report is checked.

Consider first the auxiliary loss in which the probability of a check is evaluated at the assessment rather than at the report,
\[
B[w]=\KL(Q_w\Vert P_H)+\frac{1}{2v_H}\E\bigl[\alpha(|T|)(w(T)-T)^2\bigr].
\]
For a strictly increasing $w$ with almost-everywhere derivative $w'$, the change of variables $y=w(t)$ gives
\begin{equation}\label{eq:entropy-formula}
\KL(Q_w\Vert P_H)=c_0+\E\Bigl[-\log w'(T)+\frac{w(T)^2}{2V_H}\Bigr],
\end{equation}
with $c_0$ independent of $w$. The formula holds for every strictly increasing $w$, including policies with jumps. By the change-of-variables theorem for maps that are injective and differentiable at every point of a measurable set \citep[Theorem~263D]{fremlin2016measure}, applied on the full-measure set where $w$ is differentiable, the image of $\{w'=0\}$ is Lebesgue-null. Hence both sides are infinite when $w'=0$ with positive probability, and otherwise $Q_w$ has density $f_T(t)/w'(t)$ at $w(t)$ for almost every $t$. Both $\KL(Q_w\Vert P_H)$ and $B$ are therefore convex in $w$, and the quadratic term makes them strongly convex: if $w^*$ minimizes either functional, its value at any $w$ exceeds the minimum by at least $\|w-w^*\|_2^2/(2V_H)$, where $\|w\|_2^2=\E[w(T)^2]$. By the compactness argument in the proof of \Propref{prop:policy-existence}, $B$ has a minimizer $s$. It is unique and, since $B$ is unchanged when $w(t)$ is replaced by $-w(-t)$, odd.

We claim that $|s(t)|\ge k_1|t|$ for almost every $t$. Let $G_{k_1}=\N(0,k_1^2V_L)$. By \eqref{eq:k-equation} with $a=1$, $\frac{1}{V_H}-\frac{1}{k_1^2V_L}=-\frac{k_1-1}{k_1v_H}$, so \eqref{eq:reference-gaussian} with $G=G_{k_1}$ gives, after completing the square,
\[
B[w]=\KL(Q_w\Vert G_{k_1})+c_{k_1}+\frac{1}{2k_1v_H}\E\Bigl[\alpha(|T|)(w(T)-k_1T)^2-\bigl(1-\alpha(|T|)\bigr)(k_1-1)w(T)^2\Bigr],
\]
with $c_{k_1}$ independent of $w$. Among policies with $|w(t)|\le k_1|t|$ for every $t$, each of the three terms is smallest at $w(t)=k_1t$, so $B[w]\ge B[k_1\,\mathrm{id}]$ for every such policy. Let $s_-$ and $s_+$ be the odd policies with $|s_-(t)|=\min\{|s(t)|,k_1|t|\}$ and $|s_+(t)|=\max\{|s(t)|,k_1|t|\}$. Both are strictly increasing, and at almost every $t$ their values and derivatives are those of $s$ and $k_1\,\mathrm{id}$ in some order, so \eqref{eq:entropy-formula} gives $B[s_-]+B[s_+]=B[s]+B[k_1\,\mathrm{id}]$. Since $B[s_-]\ge B[k_1\,\mathrm{id}]$, we get $B[s_+]\le B[s]$, and uniqueness gives $s_+=s$, which is the claim.

Now let $r$ be optimal for $\mathscr L$. Since $\KL(P_T\Vert P_H)=\frac{1}{2}[-\lambda-\log(1-\lambda)]\le C\lambda^2$ and $D_0\le C\lambda^2$, the comparisons $\mathscr L[r]\le\mathscr L[\mathrm{id}]$ and $B[s]\le B[\mathrm{id}]$ give $\KL(Q_w\Vert P_H)\le C\lambda^2$ for $w\in\{r,s\}$. The policy $k_0\,\mathrm{id}$ with $k_0=\sqrt{V_H/V_L}$ has report law $P_H$, so strong convexity gives $\|w-k_0\,\mathrm{id}\|_2\le C\lambda$, and since $k_0-1\le C\lambda$,
\[
\|w-\mathrm{id}\|_2\le C\lambda\qquad\text{for }w\in\{r,s\}.
\]
It follows that $\Delta_w:=\sup_t\bigl|\alpha(|w(t)|)-\alpha(|t|)\bigr|\to0$ as $\lambda\downarrow0$. Indeed, because $w$ is increasing, a discrepancy $|w(t_0)-t_0|\ge\eta$ persists with size at least $\eta/2$ on an interval of length $\eta/2$ next to $t_0$, so $w$ converges to the identity uniformly on compact sets. Outside a large compact set, $\alpha(|t|)$ is close to its limit, because $\alpha$ is bounded and weakly increasing, and so is $\alpha(|w(t)|)$, because $w$ is increasing.

By \Propref{prop:policy-relative-entropy}, the two losses differ by
\[
\mathscr L[w]-B[w]-D_0\E[\alpha(|T|)]=\E\Bigl[\bigl(\alpha(|w(T)|)-\alpha(|T|)\bigr)\Bigl(D_0+\frac{(w(T)-T)^2}{2v_H}\Bigr)\Bigr],
\]
which is at most $C\Delta_w\lambda^2$ in absolute value for $w\in\{r,s\}$. Since $\mathscr L[r]\le\mathscr L[s]$, this gives $B[r]-B[s]\le C(\Delta_r+\Delta_s)\lambda^2$, and strong convexity of $B$ gives $\|r-s\|_2^2\le C(\Delta_r+\Delta_s)\lambda^2$.

Finally, rearranging \eqref{eq:k-equation} with $a=1$ gives $(k_1-1)\bigl[\frac{1}{k_1}+\frac{V_L}{V_H}+\frac{V_L}{v_H}\bigr]=\lambda$, so $k_1-1\ge\lambda/C$. Since $|s(T)|\ge k_1|T|$, a sycophantic report at an assessment with $|T|\ge\eta$ requires $|r(T)-s(T)|\ge(k_1-1)\eta\ge\lambda\eta/C$. By Chebyshev's inequality, for every $\eta>0$,
\[
S(r)\le\Pr(|T|<\eta)+\frac{C\|r-s\|_2^2}{\lambda^2\eta^2}\le\Pr(|T|<\eta)+\frac{C(\Delta_r+\Delta_s)}{\eta^2}.
\]
The second term vanishes as $\lambda\downarrow0$, and the first vanishes as $\eta\downarrow0$. The bounds hold for every optimal policy, which implies $S(r)\to0$.
\end{proof}

\begin{proof}[Proof of \Propref{prop:uncertainty-severity}]
Label the topics so that $\beta_{L,1}>\beta_{L,2}$, hence $\chi_2>\chi_1$, and use the standardized units of \Secref{sec:uncertainty-severity}. Dividing the scaled loss $F$ of \Appref{app:surprise} by $\sigma_i^2=V_L$ and changing to standardized units gives, for each topic,
\begin{equation}\label{eq:uncertainty-normalized-loss}
\frac{F(r_i)}{\sigma_i^2}
=\frac{2v_H}{V_L}\KL\bigl(Q_{\bar r_i}\Vert\N(0,\beta_H/\beta_L)\bigr)
+\E\Bigl[\bar\alpha(|\bar r_i(X)|)\Bigl\{\frac{2v_HD_0}{V_L}+(\bar r_i(X)-X)^2\Bigr\}\Bigr],
\end{equation}
where $\beta_L$, $v_H$, $V_L$, and $D_0$ are those of topic $i$. As $\beta_H\uparrow1$, $2v_H/V_L\to0$ and $2v_HD_0/V_L\to v_L/V_L=\chi_i$. The limiting problem \eqref{eq:limit} therefore becomes the minimization of
\[
F_0^{\chi}(\bar r)=\E\bigl[\bar\alpha(|\bar r(X)|)\{\chi+(\bar r(X)-X)^2\}\bigr]
\]
with $\chi=\chi_i$, over finite nondecreasing maps in $L^2$ of the standard normal law. By \Lemref{lem:recovery}, applied in these units, minimizers exist.

We first show that the ordering holds strictly in the limit: if $\bar r_i$ minimizes $F_0^{\chi_i}$, then
\begin{equation}\label{eq:uncertainty-limit-order}
\E\bigl[(|X|-|\bar r_2(X)|)_+\bigr]>\E\bigl[(|X|-|\bar r_1(X)|)_+\bigr].
\end{equation}
As in the proof of \Lemref{lem:barrier}, projecting reports onto the interval between zero and $x$ strictly lowers the loss wherever it changes them, because $\bar\alpha(0)>0$. Every minimizer therefore lies between zero and the truth.

Suppose, toward a contradiction, that \eqref{eq:uncertainty-limit-order} fails. On positive inputs, assign the maximum of $\bar r_1,\bar r_2$ to $\chi_1$ and their minimum to $\chi_2$, leaving negative inputs unchanged. Both maps remain nondecreasing. Adding optimality inequalities cancels the distortion terms and gives
\[
(\chi_2-\chi_1)
\int_{\{x>0:\bar r_2(x)>\bar r_1(x)\}}
[\bar\alpha(\bar r_2(x))-\bar\alpha(\bar r_1(x))]\,d\Phi(x)\le0.
\]
Strict increase of $\bar\alpha$ rules out a positive-probability violation, and reflection gives the same on negative inputs, so $|\bar r_2(X)|\le|\bar r_1(X)|$ almost surely. Hence the left side of \eqref{eq:uncertainty-limit-order} is at least the right side, and since \eqref{eq:uncertainty-limit-order} fails, the two are equal. This forces $|\bar r_2|=|\bar r_1|$ almost surely, and since both lie between zero and the truth, $\bar r_1=\bar r_2$ almost surely: a single map $\bar r$ minimizes both $F_0^{\chi_1}$ and $F_0^{\chi_2}$.

This is impossible. A minimizer must have positive reports with positive probability: otherwise it is zero almost surely on positive inputs, and changing to truthful reports on a sufficiently large positive tail preserves monotonicity and lowers the loss from $\bar\alpha(0)(\chi+x^2)$ to at most $\chi$. Choose a nonnegative smooth $\psi$ supported compactly in $(0,\infty)$ with $\E[\bar\alpha'(|\bar r(X)|)\psi(\bar r(X))]>0$. Postcomposition with $y\mapsto y+\varepsilon\psi(y)$ is feasible for small $\varepsilon$ of either sign, so for $i=1,2$,
\[
0=\E\!\left[\psi(\bar r(X))\left\{
\bar\alpha'(|\bar r(X)|)[\chi_i+(\bar r(X)-X)^2]
+2\bar\alpha(|\bar r(X)|)(\bar r(X)-X)\right\}\right],
\]
where compact support and the Gaussian second moment justify differentiation. Subtracting gives $0=(\chi_2-\chi_1)\E[\bar\alpha'(|\bar r(X)|)\psi(\bar r(X))]>0$, a contradiction.

Finally, by \Lemref{lem:recovery}, applied to \eqref{eq:uncertainty-normalized-loss}, every sequence of optimal policies as $\beta_H\uparrow1$ has a subsequence along which $\bar r_i$ converges almost surely to a minimizer of $F_0^{\chi_i}$. Their severities converge by dominated convergence, since the integrands are bounded by $|X|$ even when finite-precision reports exaggerate. If the proposition failed arbitrarily close to $\beta_H=1$, choose a sequence of violating pairs and extract a common subsequence for the two topics. The limiting severity difference would be nonpositive, contradicting \eqref{eq:uncertainty-limit-order}. Thus the strict comparison holds at every sufficiently high common $\beta_H$, for all optimal choices. 
\end{proof}

\section{Increasing Verification as a Bayesian Best Response}\label{app:alpha}

This appendix characterizes when the surprise-based verification rule of \Secref{sec:model} is a Bayesian user's best response under truthful reporting. We work in the centered coordinates $\xi=\theta-\mu_0$, restrict attention to reports $d>0$ by symmetry, and recall that $v_k=(1-\beta_k)/\tau_0$ is the posterior variance of $\xi$ given a truthful report of source $k$, so that $v_H<v_L$.

\paragraph{Value of verification.} After a report $d$, the user holds a posterior over $\xi$, chooses an action $a\in\mathcal{A}$, and has utility $w(a,\xi)$; maximizers are assumed to exist with finite expected utilities. Perfect verification reveals $\xi$ before the action is chosen. The gross value of verification is the Bayesian value of information
\begin{equation*}
\mathcal{V}(d):=\E\Big[\max_{a\in\mathcal{A}}w(a,\xi)\,\Big|\,d\Big]-\max_{a\in\mathcal{A}}\E[w(a,\xi)\mid d]\ \ge0 .
\end{equation*}
Suppose the user chooses a verification intensity $q\in[0,1]$ at cost $C(q)$, with $C$ continuous on $[0,1]$, twice continuously differentiable on $(0,1)$, and $C''>0$ on $(0,1)$; with probability $q$ the state is revealed. Conditional on $d$, the user maximizes $q\,\mathcal{V}(d)-C(q)$, so the best response $\alpha^{BR}(d)$ is unique and weakly increasing in $\mathcal{V}(d)$, and at an interior solution $C'(\alpha^{BR}(d))=\mathcal{V}(d)$, whence
\begin{equation}\label{eq:alpha-br-derivative}
\alpha^{BR\prime}(d)=\frac{\mathcal{V}'(d)}{C''(\alpha^{BR}(d))} .
\end{equation}
Interior verification therefore increases with surprise exactly where the value of verification does. For quadratic target utility $w_g(a,\xi)=-\tfrac12[a-g(\xi)]^2$, where $g(\xi)$ is the full-information action, the unverified action is the posterior mean of $g(\xi)$, so $\mathcal{V}(d)=\tfrac12\operatorname{Var}(g(\xi)\mid d)$.

\paragraph{The precision effect.} Under truthful reporting, the report of source $k$ has law $\N(0,\beta_k/\tau_0)$ and $\xi\mid d\sim\N(d,v_k)$: the two sources agree on the posterior mean and differ in posterior variance. Writing $\pi_d:=\Pr(H\mid d)$,
\begin{equation*}
\log\frac{\pi_d}{1-\pi_d}=\log\frac{\pi}{1-\pi}+\half\log\frac{\bL}{\bH}+\frac{\omega}{2}d^2,\qquad
\omega:=\tau_0\Big(\frac{1}{\bL}-\frac{1}{\bH}\Big)>0,
\end{equation*}
so $\pi_d'=\omega d\,\pi_d(1-\pi_d)>0$: farther reports are more likely to have come from the advertised standard. The user's posterior variance of the state,
\begin{equation*}
V^T(d)=\pi_d v_H+(1-\pi_d)v_L,\qquad V^{T\prime}(d)=-\omega d\,\pi_d(1-\pi_d)(v_L-v_H)<0,
\end{equation*}
therefore decreases with the report's distance. Under estimation loss, $g(x)=x$, the value of verification $\tfrac12V^T$ decreases with surprise, so increasing verification requires stakes that rise with surprise.

\paragraph{Stakes versus precision.} Define the decision-stakes index $\Sigma(d):=2\mathcal{V}(d)/V^T(d)$, so that $\mathcal{V}=\half\Sigma V^T$. Under quadratic target utility, $\Sigma(d)=\operatorname{Var}(g(\xi)\mid d)/\operatorname{Var}(\xi\mid d)$, and $\Sigma\equiv1$ for $g(x)=x$. Let $\Psi(d):=-V^{T\prime}(d)/V^T(d)>0$ denote the rate at which source inference lowers residual uncertainty.

\begin{proposition}\label{prop:alpha-stakes}
Let $\Sigma$ be positive and continuously differentiable on an interval $K\subset(0,\infty)$. Then $\mathcal{V}'/\mathcal{V}=\Sigma'/\Sigma-\Psi$ on $K$, and the interior best response satisfies $\alpha^{BR\prime}(d)>0$ for every $d\in K$ if and only if
\begin{equation}\label{eq:alpha-stakes-condition}
\frac{\Sigma'(d)}{\Sigma(d)}>\Psi(d)\qquad\text{for every }d\in K .
\end{equation}
\end{proposition}

\begin{proof}
Differentiate $\log\mathcal{V}=\log\Sigma+\log V^T-\log2$ and apply \eqref{eq:alpha-br-derivative}.
\end{proof}

Increasing verification is thus a Bayesian best response exactly when the consequences of residual uncertainty grow with surprise faster than source inference removes that uncertainty. The requirement is not extreme: for $\Sigma_\eta(d)=(1+d^2)^\eta$, \Condref{eq:alpha-stakes-condition} holds for all $d>0$ whenever $\eta>\eta^\dagger:=\half\sup_{d>0}(1+d^2)\Psi(d)/d$, and the supremum is finite because $\pi_d(1-\pi_d)$ decays like $e^{-\omega d^2/2}$ while $V^T\ge v_H>0$. 

\end{appendices}
\IfFileExists{econ.bst}{\bibliographystyle{econ}}{\bibliographystyle{plainnat}}
\bibliography{bibliography}

\end{document}